\documentclass[aps,prl,twocolumn,10pt,superscriptaddress,nofootinbib,nobibnotes,longbibliography]{revtex4-1}

\usepackage{amssymb}
\usepackage{graphicx}
\usepackage{amsmath}
\usepackage{hyperref}
\usepackage{subfigure}
\usepackage{multirow}
\usepackage{setspace}
\usepackage{verbatim}
\usepackage{float}
\usepackage{color}
\usepackage{ulem}
\usepackage[utf8]{inputenc}
\usepackage[table,xcdraw]{xcolor}
\usepackage{makecell}
\usepackage{url}
\usepackage{bm}

\begin{document}

\title{Third law of repetitive electric Penrose processes}

\author{Li Hu}
\email{huli21@mails.ucas.ac.cn}
\affiliation{School of Fundamental Physics and Mathematical Sciences, Hangzhou Institute for Advanced Study (HIAS), University of Chinese Academy of Sciences (UCAS), Hangzhou 310024, China}
\affiliation{Institute of Theoretical Physics, Chinese Academy of Sciences (CAS), Beijing 100190, China}
\affiliation{University of Chinese Academy of Sciences (UCAS), Beijing 100049, China}

\author{Rong-Gen Cai}
\email{caironggen@nbu.edu.cn}
\affiliation{Institute of Fundamental Physics and Quantum Technology, \& School of Physical Science and Technology, Ningbo University, Ningbo, 315211, China}

\author{Shao-Jiang Wang}
\email{schwang@itp.ac.cn}
\affiliation{Institute of Theoretical Physics, Chinese Academy of Sciences (CAS), Beijing 100190, China}
\affiliation{Asia Pacific Center for Theoretical Physics (APCTP), Pohang 37673, Korea}

\begin{abstract}
Recently, Ruffini et al. [Phys. Rev. Lett. 134 (2025) 8, 081403] pointed out that the repetitive Penrose process cannot drain the entire extractable energy of a Kerr black hole. In this Letter, we alternatively point out the charge of a Reissner-Nordstr\"{o}m black hole cannot drop down to exactly zero via the repetitive electric Penrose process that is terminated after a finite number of iterative steps, indicating a new thermodynamical third-law analog for the repetitive electric Penrose process.
\end{abstract}
\maketitle


\textit{\textbf{Introduction.}---} 
In the original proposal of Penrose process~\cite{Penrose:1971uk}, an incoming particle decays within the ergosphere of a Kerr black hole into a negative-energy particle that falls into the black-hole horizon and a positive-energy particle that escapes to infinity.  Due to the conservation of energy, the energy of escaping particle is always greater than that of incoming particle, allowing us to extract energy from the rotating black hole. Soon after, the electric Penrose process (EPP) was proposed~\cite{Denardo:1973pyo} by replacing the Kerr background with a Reissner-Nordstr\"{o}m (RN) geometry, and finding that the negative-energy orbit also exists in the dubbed generalized ergosphere, making it feasible to extract energy from a charged black hole through the electromagnetic version of the Penrose process~\cite{Mai:2022pgf, Tursunov:2021jjf}.

Despite that the relation between the irreducible mass $M_\mathrm{irr}$ of the Penrose process and the proper horizon area $S$ has long been recognized~\cite{Christodoulou:1971pcn}, $S=16\pi M_\mathrm{irr}^2$, previous studies on iterative Penrose processes, such as black hole bombs~\cite{Feiteira:2024awb,Zaslavskii:2022vap,Kokubu:2021cwj}, usually did not update the mass and charge of the black hole after each bounce, and thus did not consider the waste in the energy extraction process. With ignorance of such waste (essentially the entropy growth), it seems feasible to extract the entire extractable energy~\footnote{For example, the total extractable energy is about $29\%$ of the total energy for an extreme Kerr black hole~\cite{Christodoulou:1971pcn}.} through the iterative Penrose process. It was only realized recently by Ruffini et al.~\cite{Ruffini:2024dwq,Ruffini:2024irc} that repeated Penrose processes for Kerr black hole of mass $M$ cannot extract all the extractable energy $E_\mathrm{extractable}\equiv M-M_\mathrm{irr}$. This is because each iteration requires updating the angular momentum and mass of a black hole with their new values, resulting in each energy extraction that actually converts a large amount of energy from $E_\mathrm{extractable}$ into a nonlinear increase in the irreducible mass $M_\mathrm{irr}$. Since the black-hole proper horizon area never decreases due to Hawking's area theorem, this part of energy will be locked up and cannot be extracted by classical processes. 

In this Letter, we turn to the repetitive EPP in the RN black hole. We find a similar incapability of extracting all extractable energy, and the charge of the RN black hole can be very small eventually, but never vanishes solely from the EPP that absorbs a decay product particle with negative energy. The charge of an RN black hole can only possibly vanish either from the semiclassical/quantum process by Hawking evaporation~\cite{Brown:2024ajk} or accidental neutralization by absorbing positive-energy particles with exactly equal but opposite charge.

\textit{\textbf{Electric Penrose process.}---}
The metric of an RN black hole of mass $M$ and charge $Q$ can be described as
\begin{align}
    \mathrm{d}s^2=-f\mathrm{d}t^2+f^{-1}\mathrm{d}r^2+r^2\mathrm{d}\theta^2+r^2\sin^2\theta\mathrm{d}\phi^2\,,
\end{align}
where $f=1-\frac{2M}{r}+\frac{Q^2}{r^2}$. Then, we consider EPP with an incident particle 0 from infinity decaying at $r_d$, producing a negative-energy particle 1 that falls into the black hole and a positive-energy particle 2 that escapes to infinity. The four-momentum and charge conservation laws read
\begin{align}
    \hat{E}_0&=\tilde{\mu}_1\hat{E}_1+\tilde{\mu}_2\hat{E}_2\,,\label{eq:E0}\\
    \hat{q}_0&=\tilde{\mu}_1\hat{q}_1+\tilde{\mu}_2\hat{q}_2\,,\label{eq:q0}\\
    \hat{P}_0&=\tilde{\mu}_1\hat{P}_1+\tilde{\mu}_2\hat{P}_2\,,\label{eq:P0}
\end{align}
where $\hat{E}_i=E_i/\mu_i$, $\hat{q}_i=q_i/\mu_i$, and $\hat{P}_i=P_i/\mu_i$ represent the dimensionless energy, charge, and radial momentum, respectively, $\mu_i$ is the mass of particle $i$, and $\tilde{\mu}_i=\mu_i/\mu_0$.

Recall that in Ref.~\cite{Zaslavskii:2024zgh} the decay products are ejected along the same trajectory of the incident particle to achieve a maximum efficiency. Since we are dealing with the RN metric, we further simplify the configuration of the incident particle from infinity, either at rest or with an initial radial velocity~\cite{Kokubu:2021cwj}, so that both particles 1 and 2 maintain radial motion to maximize the energy return on investment (EROI) (defined later).

Besides the conservation Eqs.~\eqref{eq:E0}, \eqref{eq:q0}, and~\eqref{eq:P0}, the normalization of the proper four velocity gives rise to another three equations,
\begin{align}\label{eq:Potential}
    \hat{P}_i^2&=(\hat{E}_i-\hat{V}_i^{-})(\hat{E}_i-\hat{V}_i^{+})\,,\quad \hat{Q}\equiv Q/M\nonumber\\
    &=(\hat{E}_i-\frac{\hat{Q}\hat{q}_i}{\hat{r}}+\sqrt{f})(\hat{E}_i-\frac{\hat{Q}\hat{q}_i}{\hat{r}}-\sqrt{f})\,,
\end{align}
for $i=0,1,2$, where $\hat{V}_i^{\pm}$ is understood as the effective potential of the radial motion~\cite{Ruffini:1971bza}, and $\hat{r}\equiv r/M$ is the dimensionless radius. Now we have six equations and 11 variables (two masses, three energies, three charges, three radial momenta). However, the maximum EROI condition can further eliminate  Eq.~\eqref{eq:P0}. References~\cite{Ruffini:2024dwq,Ruffini:2024irc} show that in the Penrose process, the maximum EROI corresponds to the decay when all three particles are at their own turning points~\cite{Ruffini:2024irc} with vanishing kinetic energy, that is, the radial momenta of the three particles must vanish at the decay location. We will show shortly below that this correspondence still holds for EPP.

\textit{\textbf{Maximum EROI.}---}
At the decay position, since $\hat{E}_1$ is supposed to be negative for any Penrose process, we require $\hat{E}_1$ to be as negative as possible for given reduced energy $\hat{E}_0$ to maximize the single EROI $\xi\equiv E_2/E_0-1=-\tilde{\mu}_1\hat{E}_1/\hat{E}_0$. The lower limit is at its turning point with $\hat{E}_1=\hat{V}_1^{+}$; exceeding this limit is the classical forbidden region. According to the conservation of radial momenta, we have $P_0=P_2$ due to $P_1=0$. To further prove that both the radial momenta $P_0=P_2$ vanish, let us first assume them to be nonzero, then Eq.~\eqref{eq:Potential} reduces to
\begin{equation}
\left\{
\begin{aligned}
    E_0-V_0^{-}&=E_2-V_2^{-}\,,\\
    E_0-V_0^{+}&=E_2-V_2^{+}\,.
\end{aligned}
\right.
\end{equation}
Combining the above two equations, we have
\begin{align}
    E_0-\frac{\hat{Q}q_0}{\hat{r}}=E_2-\frac{\hat{Q}q_2}{\hat{r}}\,.
\end{align}
The conservation of energy and charge further leads to
\begin{align}
    \hat{E}_1=\frac{\hat{Q}\hat{q}_1}{\hat{r}}\,,
\end{align}
resulting in a violation of the timelike geodesic condition for particle 1,
\begin{align}
    \Big(\frac{\mathrm{d}t}{\mathrm{d}\tau}\Big)_1=\frac{\hat{E}_1-\hat{Q}\hat{q}_1/\hat{r}}{\sqrt{f}}>0\,,
\end{align}
where $\tau$ is the proper time along the geodesic of particle 1. Therefore, all three particles must be located at their turning points with vanishing kinetic energy at the decay location to maximize the efficiency, and hence eliminating Eq.~\eqref{eq:P0}.

Finally, we obtain five equations~\eqref{eq:E0}, \eqref{eq:q0}, and~\eqref{eq:Potential}, and eight unknown variables $\tilde{\mu}_{1,2}$, $\hat{E}_{0,1,2}$, and $\hat{q}_{0,1,2}$. We choose to vary $\hat{E}_0$, $\hat{q}_1$ for given $\eta=\mu_2/\mu_1$, and use them to calculate the other five variables. The solutions are
\begin{align}
    \tilde{\mu}_2&=\eta\,\tilde{\mu}_1=\frac{\eta}{1+\eta}\,,\\
    \hat{E}_1&=\frac{\hat{Q}\hat{q}_1}{\hat{r}_d}+\sqrt{1-\frac{2}{\hat{r}_d}+\frac{\hat{Q}^2}{\hat{r}_d^2}}\,,\\
    \hat{q}_0&=\frac{\hat{r}_d}{\hat{Q}}\Big(\hat{E}_0-\sqrt{1-\frac{2}{\hat{r}_d}+\frac{\hat{Q}^2}{\hat{r}_d^2}}\Big)\,,\\
    \hat{q}_2&=\frac{\hat{q}_0-\tilde{\mu}_1\hat{q}_1}{\tilde{\mu}_2}\,,\\
    \hat{E}_2&=\frac{\hat{Q}\hat{q}_2}{\hat{r}_d}+\sqrt{1-\frac{2}{\hat{r}_d}+\frac{\hat{Q}^2}{\hat{r}_d^2}}\,,
\end{align}
where $\hat{r}_d$ denotes the dimensionless decay radius.

\textit{\textbf{Repetitive stop conditions.}---}
After each energy extraction, the remaining mass and charge of the black hole are
\begin{align}
    M_n&=M_{n-1}+\mu_0\tilde{\mu}_1\hat{E}_{1,n-1}\,,\\
    Q_n&=Q_{n-1}+\mu_0\tilde{\mu}_1\hat{q}_1\,.
\end{align}
To determine the condition that terminates this iteration, note that it is influenced by the following two factors. 

First, particle 1 must fall into the black hole, while particles 0 and 2 must escape to infinity. For the negatively charged particle 1, $\hat{V}_1^+(\hat{r})$ is monotonically increasing, ensuring it always crosses the event horizon. For particles 0 and 2 to escape, their classical turning points must lie outside the peak of their respective effective potentials. The limiting condition for each is that its turning point coincides with its potential peak, satisfying $\hat{V}_i^+(\hat{r}_d)=\hat{E}_i$, and $\mathrm{d}\hat{V}_i^+/\mathrm{d}\hat{r}|_{\hat{r}=\hat{r}_d}=0$. Solving them yields a minimal $\hat{Q}$ for each particle,
\begin{align}
    \hat{Q}^2_{i,\mathrm{min}}=\frac{(\hat{r}_d-1)^2}{\hat{E}_i^2}+\hat{r}_d(2-\hat{r}_d)\,,\quad i=0,2\,.
\end{align}
Since $E_2 > E_0$ implies $\hat{E}_2 > \hat{E}_0$, the constraint from particle 2 is weaker. Thus, the lower limit on $\hat{Q}$ is actually determined by particle 0,
\begin{align}\label{eq:con1}
    \hat{Q}_{\mathrm{min}}^2=\frac{(\hat{r}_d-1)^2}{\hat{E}_0^2}+\hat{r}_d(2-\hat{r}_d)\,.
\end{align}
It is worth noting that if the particle 0 is released from rest at infinity ($\hat{E}_0=1$), $\hat{Q}_{\mathrm{min}}$ is always unity, rendering this initial configuration forbidden in EPP.


Second, $\hat{E}_1$ generated at each iteration must be strictly negative, which requires
\begin{equation}
\left\{
\begin{aligned}
    \hat{Q}&>\sqrt{\frac{\hat{r}(\hat{r}-2)}{\hat{q}_1^2-1}},\quad |\hat{q}_1|>1;\\
    \hat{Q}&<\sqrt{\frac{\hat{r}(\hat{r}-2)}{\hat{q}_1^2-1}},\quad 0<|\hat{q}_1|<1.
\end{aligned}
\right.
\end{equation}

\textit{\textbf{The third-law analog.}---}
When $\hat{E}_0$ and $\hat{q}_1$ are given, for the electric Penrose process to successfully occur, the choice of the location where the decay occurs and the charge of the black hole are severely restricted. 

\begin{figure}[h]
    \centering
    \includegraphics[width=0.48\textwidth]{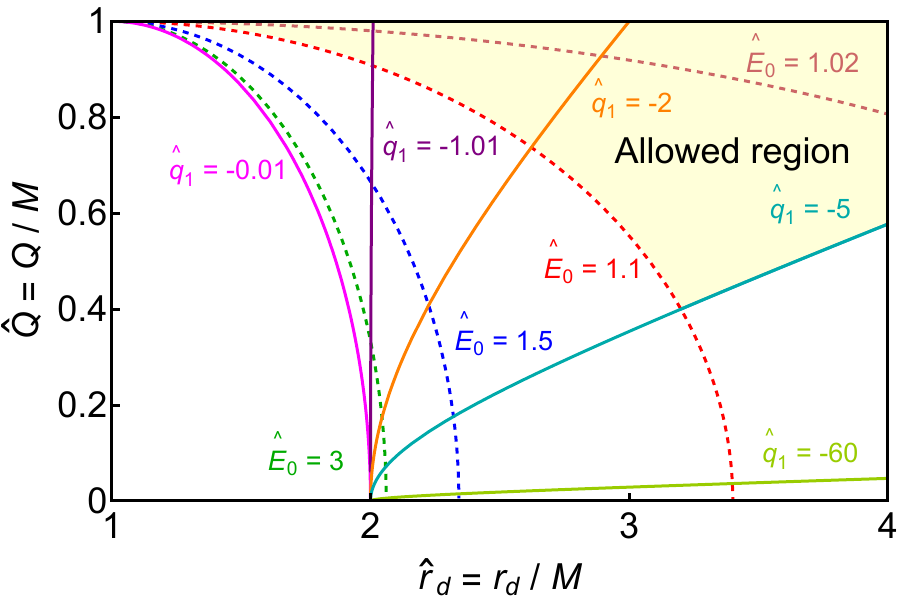}
	\caption{Allowed region of the electric Penrose process derived from two iterative stopping conditions for various values of $\hat{E}_0$ and $\hat{q}_1$. The allowed region (shaded region) is to the left of the solid curve and to the right of the dashed curve.}
    \label{fig:Allowed_region}
\end{figure}

In Fig.~\ref{fig:Allowed_region}, we show the restrictions in the parameter space of $\hat{E}_0$ and $\hat{q}_1$. For a given pair of $\hat{E}_0$ and $\hat{q}_1$, two corresponding limiting curves (dashed and solid) are obtained. The intersection region between the right-hand side of the dashed curve and the left-hand side of the solid curve is the allowed region. After that, a decay location $\hat{r}_d$ is chosen in the allowed region, and the black hole is then evolved from upper (larger charge) to lower (smaller charge). Note that, when $-1<\hat{q}_1<0$ (macroscopic object with small charge-to-mass ratio), the allowed region is below the solid curve, indicating that EPP cannot always start from an extreme RN black hole.

Typically, the charge-to-mass ratio of microscopic particles is huge, hence the lower limit $\hat{Q}_\mathrm{min}$ of the reduced charge (solid curve) can be extremely close to zero but not exactly zero. Meanwhile, there are a large number of high-energy particles $\hat{E}_0>1$ in the Universe that can serve as incident particles. This means that the EPP occurring in the allowed region of the generalized ergosphere at $\hat{r}>2$ will reduce the charge of the RN black hole to be arbitrarily close to zero but not vanishing.

\textit{\textbf{Energy extraction efficiencies.}---} 
For an RN black hole, the irreducible mass of $n$-th repetitive EPP is~\cite{Christodoulou:1971pcn}
\begin{align}
    M_{\mathrm{irr},n}=\frac{M_n+\sqrt{M_n^2-Q_n^2}}{2}\,,
\end{align}
which is nonlinear in $M_n$ and $Q_n$. The corresponding total extractable energy $E_{\mathrm{extractable},n}=M_n-M_{\mathrm{irr},n}$ is
\begin{align}
    E_{\mathrm{extractable},n}=\frac{M_n-\sqrt{M_n^2-Q_n^2}}{2}\,.
\end{align}
For an extreme charged black hole with $Q/M=1$, half of the total energy is extractable. However, we will see that we cannot extract all of them.

To measure the maximum efficiency of energy extraction, we define the EROI as the energy ratio of all product particles eventually falling into the black hole with respect to all incident particles from infinity, $\xi_n=-\sum_{j=1}^n E_{1,\,j-1}/(nE_0)\equiv E_{\mathrm{extracted},n}/(n E_0)$, which is found numerically easy to exceed $100\%$. However, the energy utilization efficiency (EUE), defined as an energy ratio of all product particles eventually falling into the black hole with respect to the difference between the initial and final extractable energy, $\Xi_n=E_{\mathrm{extracted},n}/(E_{\mathrm{extractable},0}-E_{\mathrm{extractable},n})\equiv E_{\mathrm{extracted},n}/
\Delta E_{\mathrm{extractable},n}$, which is found numerically difficult to exceed $50\%$, not to mention to extract all the extractable energy of an extremely charged black hole.

\begin{table}[htbp]
\caption{Parameters and initial conditions selected for repetitive electric Penrose process. Here we use the subscript ``able'' to represent ``extractable'' and ``ed'' to represent ``extracted''. We take $\mu_0=M_0/100$ to increase the mass and charge extracted through a single EPP, thereby reducing the number of iterations required for this demonstration. Of course, one can also take a much smaller $\mu_0$, which requires more iterations. We also set the decay to occur at $\hat{r}=2.4$, and the mass ratio $\eta=0.78345$ is taken from Ref.~\cite{Ruffini:2024irc}. The initial energy of particle 0 at infinity is chosen to be $\hat{E}_0=1.1$, and the charge is $\hat{q}_1=-5$, which are taken from the example in Fig.~\ref{fig:Allowed_region}. With the iterative stop condition, the lower limit of the reduced charge is obtained, $\hat{Q}_{\mathrm{min}}\approx 0.8123$.}
\label{tab:1}
\centering
\renewcommand{\arraystretch}{1.5}
\setlength{\tabcolsep}{4pt}
{\scriptsize
\begin{tabular}{cccccccc}
\hline
\hline\\[-10pt]
$n$ & $\hat{Q}_n$ & $\frac{M_n}{M_0}$ & $\frac{E_{\mathrm{able},n}}{M_0}$ & $\frac{E_{\mathrm{ed},n}}{M_0}$ & $\frac{M_{\mathrm{irr},n}}{M_0}$ & $\xi_n$ & $\Xi_n$\\[1.5pt]
\hline
0 & 1.0000 & 1.0000 & 0.5000 & 0.0000 & 0.5000 & 0.0000 & 0.0000\\
1 & 0.9802 & 0.9916 & 0.3976 & 0.0084 & 0.5939 & 0.7646 & 0.0822\\
2 & 0.9599 & 0.9834 & 0.3538 & 0.0166 & 0.6296 & 0.7556 & 0.1137\\
3 & 0.9390 & 0.9754 & 0.3200 & 0.0246 & 0.6554 & 0.7464 & 0.1368\\
4 & 0.9176 & 0.9676 & 0.2915 & 0.0324 & 0.6761 & 0.7370 & 0.1555\\
5 & 0.8957 & 0.9600 & 0.2665 & 0.0400 & 0.6935 & 0.7275 & 0.1714\\
6 & 0.8732 & 0.9526 & 0.2441 & 0.0474 & 0.7085 & 0.7178 & 0.1852\\
7 & 0.8501 & 0.9455 & 0.2238 & 0.0545 & 0.7217 & 0.7079 & 0.1973\\
8 & 0.8265 & 0.9386 & 0.2051 & 0.0614 & 0.7335 & 0.6978 & 0.2082\\
9 & 0.8023 & 0.9319 & 0.1878 & 0.0681 & 0.7441 & 0.6875 & 0.2180\\
\hline
\end{tabular}
}
\end{table}

In Tab.~\ref{tab:1}, we present an example of the complete iterative process corresponding to the parameter choices of the light yellow shaded region in Fig.~\ref{fig:Allowed_region} ($\hat{E}_0=1.1\,,\hat{q}_1=-5$) and set the decay position to $\hat{r}_d=2.4$. In this scenario, the lower limit of the reduced charge is $\hat{Q}_{\mathrm{min}}\approx 0.8123$. Besides, we also choose the mass of particle 0 to $\mu_0=M_0/100$. Of course, there are generally no single particles with such a large mass. The value we choose here is just to speed up the iteration process in this demonstration. It is straightforward to find that the iteration triggers the stopping condition after nine steps, and the EROI of the repetitive process is about $69\%$, which is much higher than the original proposal of the Penrose process.

\textit{\textbf{Repetitive electric Penrose process.}---} 
Looking more carefully into Table~\ref{tab:1}, one can find that between two arbitrary steps, the decrease in extractable energy is channeled into both the extracted energy and the irreducible mass, due to the nonlinear behavior of the latter,
\begin{align}
    E_{\mathrm{extractable},n}|_i^j= E_{\mathrm{extracted},n}|_i^j+ M_{\mathrm{irr},n}|_i^j\,.
\end{align}
Consequently, the overall EUE stands at only $22\%$, with nearly three-quarters of the reduced extractable energy being converted into the irreducible mass of the black hole (that is, the increase in entropy). This portion is thereby permanently locked up according to Hawking's area law, imposing a fundamental thermodynamic cost on the energy extraction process (see also Ref.~\cite{Bravetti:2015xsp}).

\begin{figure}[h]
    \centering
    \includegraphics[width=0.48\textwidth]{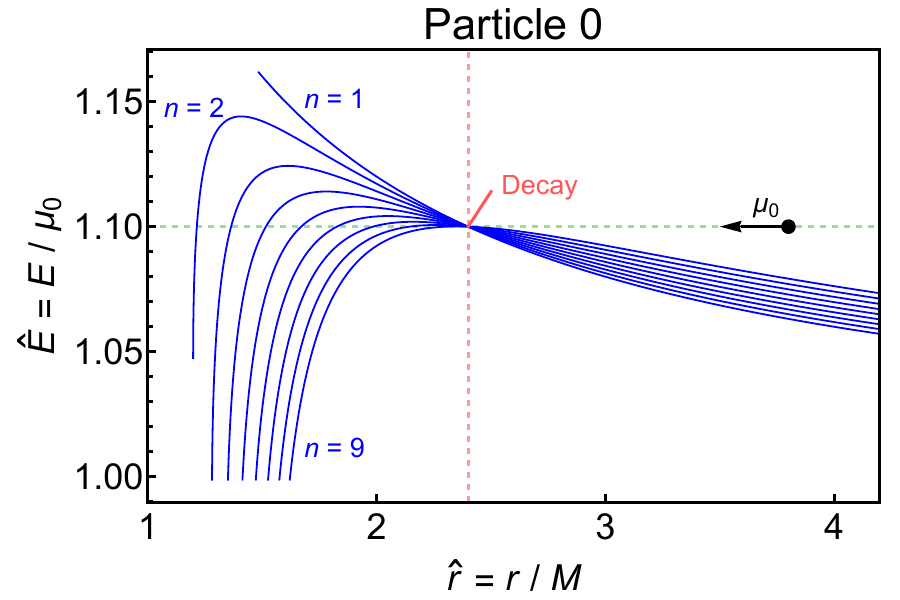}
	\caption{Schematic diagram of particle 0 decay. Parameters are the same as Table~\ref{tab:1}. The red dashed line indicates the decay position at $\hat{r}=2.4$, the green dashed line indicates the incident energy $\hat{E}_0$, and the blue solid curves represent the effective potential at each iteration. It can be seen that the decay position is always to the right of the peak of the effective potential.}
    \label{fig:Particle_0}
\end{figure}

We also provide Fig.~\ref{fig:Particle_0} to show the change of the effective potential of the incident particle during the iteration. One can find that the peak value of the effective potential decreases as the number of iterations increases. Before the ninth iteration, although the peak of the effective potential is still to the left of the decay position, its peak value is very close to the incident energy, which clearly shows how the iteration stops due to Eq.~\eqref{eq:con1}.


\begin{figure*}[htbp]
    \centering
    \includegraphics[width=1\textwidth]{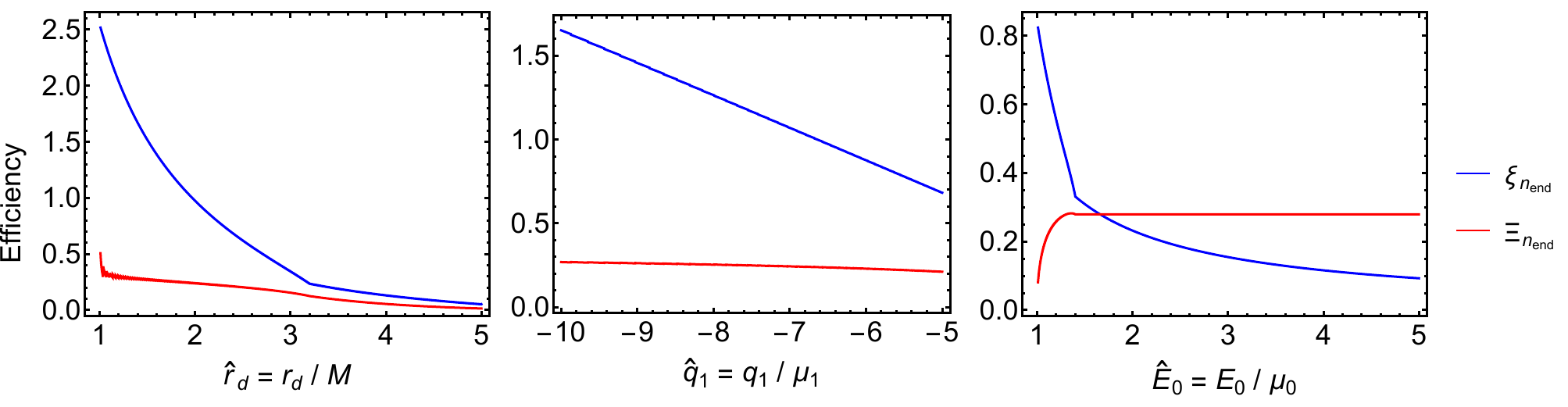}
	\caption{Variation of EROI and EUE of the final state with the dimensionless decay radius, the charge-to-mass ratio of particle 1, and the reduced energy of particle 0. Here we keep the mass ratio $\eta=0.78345$ but choose a smaller mass of the incident particle $\mu_0=M_0/1000$ to eliminate the ``oscillation" of the curves as much as possible. The basic choice of three variables is $\hat{r}_d=2.4$, $\hat{q}_1=-5$, and $\hat{E}_0=1.1$. Each time one of the three variables is changed while keeping the other two fixed. It is intuitive to see that $\xi_{n_\mathrm{end}}$ can easily be much greater than $1$, but $\Xi_{n_\mathrm{end}}$ is difficult to exceed $0.5$.}
    \label{fig:Efficiency}
\end{figure*}

Furthermore, we also vary the dimensionless decay radius $\hat{r}_d$, the charge-to-mass ratio $\hat{q}_1$ of particle 1, and the reduced energy $\hat{E}_0$ of particle 0, and study the final EROI and EUE of the repetitive process from the beginning of an extreme black hole to the end of iteration. The results are presented in Fig.~\ref{fig:Efficiency}.

It is straightforward to find that both EROI and EUE decrease significantly as the decay location moves farther away from the black hole horizon, while both of them increase as the charge-to-mass ratio of particle 1 increases. Regarding the velocity of the incident particle, since the energy extracted by a single process is nearly ``fixed'' by $\hat{q}_1$, EROI decreases as $\hat{E}_0$ becomes bigger. Interestingly, EUE stops growing and becomes constant when $\hat{E}_0$ exceeds $1.4$. This is because the stopping condition switches from the dashed curve in Fig.~\ref{fig:Allowed_region} to the solid curve when $\hat{E}_0>1.4$, that is, the final state of the black hole is fixed and will no longer change with $\hat{E}_0$. One thing to note is that we do not show the case $|\hat{q}_1|<1$ here because it may not be an extreme black hole initially, making it difficult to compare it with other cases directly.


\textit{\textbf{Conclusions and discussions.}---} 
We look into the repetitive EPP under the condition of maximizing the EROI. We first find the parameter choices (the dimensionless decay position and reduced BH charge) are constrained by both the incident particle velocity and the charge-to-mass ratio of the product particle that falls into the black hole. Given the substantially high charge-to-mass ratio of microscopic particles and the abundance of high-energy cosmic particles in the Universe, the charge of the RN black hole can be efficiently neutralized to a negligibly small value through repetitive EPP. Nevertheless, it is noteworthy that the charge cannot be fully discharged to zero via this classical process, revealing another fundamental limit in a classic BH process.

Furthermore, we demonstrate that, just as reducing a black-hole spin does not amount to extracting all the corresponding rotational energy, so does reducing its charge fail to extract all the corresponding extractable electrical energy---a fundamental limitation rooted in the nonlinear growth of the irreducible mass.

In addition, we stress that EROI can easily exceed $100\%$ and hence it is feasible to obtain large returns with small inputs in repetitive EPP. Unfortunately, the EUE can hardly exceed $50\%$, so that most of the extractable energy is wasted in repetitive EPP. Admittedly, increasing the charge-to-mass ratio of particle 1, decaying closer to the black hole, and appropriately increasing the speed of the incident particle can further increase EUE, but only at the cost of significant fine-tuning.

For future work, a critical direction is to model the concurrent evolution of charge and spin for a Kerr-Newman black hole under repetitive EPP. Furthermore, extending this study to repetitive superradiance, the wave counterpart of this energy extraction mechanism, would also be highly illuminating.

\textit{\textbf{Acknowledgments.}---} 
This work is supported by the National Key Research and Development Program of China Grants No. 2021YFC2203004, No. 2021YFA0718304, and No. 2020YFC2201501, the National Natural Science Foundation of China Grants No. 12422502, No. 12547110, No. 12588101, No. 12235019, and No. 12447101, and the China Manned Space Program Grant No. CMS-CSST-2025-A01.

\textit{\textbf{Data availability.}---} No data were created or analyzed in this study.

\section{End matter}

\subsection{Improvement to previous work}

\begin{figure}[htbp]
    \centering
    \includegraphics[width=0.48\textwidth]{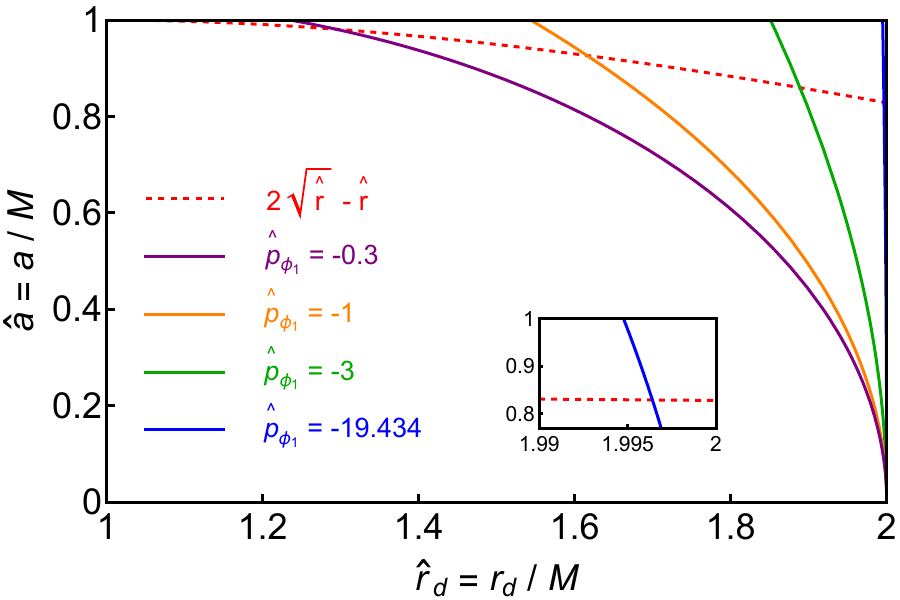}
	\caption{Allowed region of the original Penrose process derived from the iterative stopping condition, which is the overlapping region to the upper right of the red dashed curve (previous constraint) and to the lower left of a solid curve (newly added constraint). The parameter of the blue curve $\hat{p}_{\phi1}=-19.434$ is taken from~\cite{Ruffini:2024dwq}, indicating that the condition $\hat{E}_1<0$ must be considered.}
    \label{fig:Kerr}
\end{figure}

The constraint obtained in Ruffini et al.~\cite{Ruffini:2024dwq,Ruffini:2024irc} can be further improved by including the second stopping condition (it might be more accurate to call it the starting condition here) $\hat{E}_1<0$, rendering $\hat{a}<\sqrt{\frac{2\hat{p}_{\phi1}^2-\hat{p}_{\phi1}^2\hat{r}^2+2\hat{r}^2-\hat{r}^3}{\hat{r}}}$ on the equatorial plane, which has further shrunk the allowed region in the parameter space as shown in Fig.~\ref{fig:Kerr}. The spin parameter $\hat{a}$ is defined in terms of the angular momentum $L$ as $\hat{a}\equiv a/M=L/M^2$, while the dimensionless angular momentum of particle 1 is given by $\hat{p}_{\phi1}\equiv p_{\phi1}/(\mu_1 M)$.

\bibliography{ref}

\begin{thebibliography}{14}%
\makeatletter
\providecommand \@ifxundefined [1]{%
 \@ifx{#1\undefined}
}%
\providecommand \@ifnum [1]{%
 \ifnum #1\expandafter \@firstoftwo
 \else \expandafter \@secondoftwo
 \fi
}%
\providecommand \@ifx [1]{%
 \ifx #1\expandafter \@firstoftwo
 \else \expandafter \@secondoftwo
 \fi
}%
\providecommand \natexlab [1]{#1}%
\providecommand \enquote  [1]{``#1''}%
\providecommand \bibnamefont  [1]{#1}%
\providecommand \bibfnamefont [1]{#1}%
\providecommand \citenamefont [1]{#1}%
\providecommand \href@noop [0]{\@secondoftwo}%
\providecommand \href [0]{\begingroup \@sanitize@url \@href}%
\providecommand \@href[1]{\@@startlink{#1}\@@href}%
\providecommand \@@href[1]{\endgroup#1\@@endlink}%
\providecommand \@sanitize@url [0]{\catcode `\\12\catcode `\$12\catcode
  `\&12\catcode `\#12\catcode `\^12\catcode `\_12\catcode `\%12\relax}%
\providecommand \@@startlink[1]{}%
\providecommand \@@endlink[0]{}%
\providecommand \url  [0]{\begingroup\@sanitize@url \@url }%
\providecommand \@url [1]{\endgroup\@href {#1}{\urlprefix }}%
\providecommand \urlprefix  [0]{URL }%
\providecommand \Eprint [0]{\href }%
\providecommand \doibase [0]{http://dx.doi.org/}%
\providecommand \selectlanguage [0]{\@gobble}%
\providecommand \bibinfo  [0]{\@secondoftwo}%
\providecommand \bibfield  [0]{\@secondoftwo}%
\providecommand \translation [1]{[#1]}%
\providecommand \BibitemOpen [0]{}%
\providecommand \bibitemStop [0]{}%
\providecommand \bibitemNoStop [0]{.\EOS\space}%
\providecommand \EOS [0]{\spacefactor3000\relax}%
\providecommand \BibitemShut  [1]{\csname bibitem#1\endcsname}%
\let\auto@bib@innerbib\@empty
\bibitem [{\citenamefont {Penrose}\ and\ \citenamefont
  {Floyd}(1971)}]{Penrose:1971uk}%
  \BibitemOpen
  \bibfield  {author} {\bibinfo {author} {\bibfnamefont {R.}~\bibnamefont
  {Penrose}}\ and\ \bibinfo {author} {\bibfnamefont {R.~M.}\ \bibnamefont
  {Floyd}},\ }\bibfield  {title} {\enquote {\bibinfo {title} {{Extraction of
  rotational energy from a black hole}},}\ }\href {\doibase
  10.1038/physci229177a0} {\bibfield  {journal} {\bibinfo  {journal} {Nature}\
  }\textbf {\bibinfo {volume} {229}},\ \bibinfo {pages} {177--179} (\bibinfo
  {year} {1971})}\BibitemShut {NoStop}%
\bibitem [{\citenamefont {Denardo}\ and\ \citenamefont
  {Ruffini}(1973)}]{Denardo:1973pyo}%
  \BibitemOpen
  \bibfield  {author} {\bibinfo {author} {\bibfnamefont {G.}~\bibnamefont
  {Denardo}}\ and\ \bibinfo {author} {\bibfnamefont {R.}~\bibnamefont
  {Ruffini}},\ }\bibfield  {title} {\enquote {\bibinfo {title} {{On the
  energetics of Reissner Nordstr{\o}m geometries}},}\ }\href {\doibase
  10.1016/0370-2693(73)90198-6} {\bibfield  {journal} {\bibinfo  {journal}
  {Phys. Lett. B}\ }\textbf {\bibinfo {volume} {45}},\ \bibinfo {pages}
  {259--262} (\bibinfo {year} {1973})}\BibitemShut {NoStop}%
\bibitem [{\citenamefont {Mai}\ and\ \citenamefont {Yang}(2023)}]{Mai:2022pgf}%
  \BibitemOpen
  \bibfield  {author} {\bibinfo {author} {\bibfnamefont {Zhan-Feng}\
  \bibnamefont {Mai}}\ and\ \bibinfo {author} {\bibfnamefont {Run-Qiu}\
  \bibnamefont {Yang}},\ }\bibfield  {title} {\enquote {\bibinfo {title}
  {{Black holes as rechargeable batteries and nuclear reactors}},}\ }\href
  {\doibase 10.1103/PhysRevD.108.104066} {\bibfield  {journal} {\bibinfo
  {journal} {Phys. Rev. D}\ }\textbf {\bibinfo {volume} {108}},\ \bibinfo
  {pages} {104066} (\bibinfo {year} {2023})},\ \Eprint
  {http://arxiv.org/abs/2210.10587} {arXiv:2210.10587 [gr-qc]} \BibitemShut
  {NoStop}%
\bibitem [{\citenamefont {Tursunov}\ \emph {et~al.}(2021)\citenamefont
  {Tursunov}, \citenamefont {Juraev}, \citenamefont {Stuchl{\'\i}k},\ and\
  \citenamefont {Kolo{\v{s}}}}]{Tursunov:2021jjf}%
  \BibitemOpen
  \bibfield  {author} {\bibinfo {author} {\bibfnamefont {Arman}\ \bibnamefont
  {Tursunov}}, \bibinfo {author} {\bibfnamefont {Bakhtinur}\ \bibnamefont
  {Juraev}}, \bibinfo {author} {\bibfnamefont {Zden{\v{e}}k}\ \bibnamefont
  {Stuchl{\'\i}k}}, \ and\ \bibinfo {author} {\bibfnamefont {Martin}\
  \bibnamefont {Kolo{\v{s}}}},\ }\bibfield  {title} {\enquote {\bibinfo {title}
  {{Electric Penrose process: High-energy acceleration of ionized particles by
  nonrotating weakly charged black hole}},}\ }\href {\doibase
  10.1103/PhysRevD.104.084099} {\bibfield  {journal} {\bibinfo  {journal}
  {Phys. Rev. D}\ }\textbf {\bibinfo {volume} {104}},\ \bibinfo {pages}
  {084099} (\bibinfo {year} {2021})},\ \Eprint
  {http://arxiv.org/abs/2109.10288} {arXiv:2109.10288 [gr-qc]} \BibitemShut
  {NoStop}%
\bibitem [{\citenamefont {Christodoulou}\ and\ \citenamefont
  {Ruffini}(1971)}]{Christodoulou:1971pcn}%
  \BibitemOpen
  \bibfield  {author} {\bibinfo {author} {\bibfnamefont {D.}~\bibnamefont
  {Christodoulou}}\ and\ \bibinfo {author} {\bibfnamefont {R.}~\bibnamefont
  {Ruffini}},\ }\bibfield  {title} {\enquote {\bibinfo {title} {{Reversible
  transformations of a charged black hole}},}\ }\href {\doibase
  10.1103/PhysRevD.4.3552} {\bibfield  {journal} {\bibinfo  {journal} {Phys.
  Rev. D}\ }\textbf {\bibinfo {volume} {4}},\ \bibinfo {pages} {3552--3555}
  (\bibinfo {year} {1971})}\BibitemShut {NoStop}%
\bibitem [{\citenamefont {Feiteira}\ \emph {et~al.}(2024)\citenamefont
  {Feiteira}, \citenamefont {Lemos},\ and\ \citenamefont
  {Zaslavskii}}]{Feiteira:2024awb}%
  \BibitemOpen
  \bibfield  {author} {\bibinfo {author} {\bibfnamefont {Duarte}\ \bibnamefont
  {Feiteira}}, \bibinfo {author} {\bibfnamefont {Jos{\'e} P.~S.}\ \bibnamefont
  {Lemos}}, \ and\ \bibinfo {author} {\bibfnamefont {Oleg~B.}\ \bibnamefont
  {Zaslavskii}},\ }\bibfield  {title} {\enquote {\bibinfo {title} {{Penrose
  process in Reissner-Nordstr{\"o}m-AdS black hole spacetimes: Black hole
  energy factories and black hole bombs}},}\ }\href {\doibase
  10.1103/PhysRevD.109.064065} {\bibfield  {journal} {\bibinfo  {journal}
  {Phys. Rev. D}\ }\textbf {\bibinfo {volume} {109}},\ \bibinfo {pages}
  {064065} (\bibinfo {year} {2024})},\ \Eprint
  {http://arxiv.org/abs/2401.13039} {arXiv:2401.13039 [gr-qc]} \BibitemShut
  {NoStop}%
\bibitem [{\citenamefont {Zaslavskii}(2022)}]{Zaslavskii:2022vap}%
  \BibitemOpen
  \bibfield  {author} {\bibinfo {author} {\bibfnamefont {O.~B.}\ \bibnamefont
  {Zaslavskii}},\ }\bibfield  {title} {\enquote {\bibinfo {title} {{Confined
  Penrose process and black-hole bomb}},}\ }\href {\doibase
  10.1103/PhysRevD.106.024037} {\bibfield  {journal} {\bibinfo  {journal}
  {Phys. Rev. D}\ }\textbf {\bibinfo {volume} {106}},\ \bibinfo {pages}
  {024037} (\bibinfo {year} {2022})},\ \Eprint
  {http://arxiv.org/abs/2204.12405} {arXiv:2204.12405 [gr-qc]} \BibitemShut
  {NoStop}%
\bibitem [{\citenamefont {Kokubu}\ \emph {et~al.}(2021)\citenamefont {Kokubu},
  \citenamefont {Li}, \citenamefont {Wu},\ and\ \citenamefont
  {Yu}}]{Kokubu:2021cwj}%
  \BibitemOpen
  \bibfield  {author} {\bibinfo {author} {\bibfnamefont {Takafumi}\
  \bibnamefont {Kokubu}}, \bibinfo {author} {\bibfnamefont {Shou-Long}\
  \bibnamefont {Li}}, \bibinfo {author} {\bibfnamefont {Puxun}\ \bibnamefont
  {Wu}}, \ and\ \bibinfo {author} {\bibfnamefont {Hongwei}\ \bibnamefont
  {Yu}},\ }\bibfield  {title} {\enquote {\bibinfo {title} {{Confined Penrose
  process with charged particles}},}\ }\href {\doibase
  10.1103/PhysRevD.104.104047} {\bibfield  {journal} {\bibinfo  {journal}
  {Phys. Rev. D}\ }\textbf {\bibinfo {volume} {104}},\ \bibinfo {pages}
  {104047} (\bibinfo {year} {2021})},\ \Eprint
  {http://arxiv.org/abs/2108.13768} {arXiv:2108.13768 [gr-qc]} \BibitemShut
  {NoStop}%
\bibitem [{\citenamefont {Ruffini}\ \emph
  {et~al.}(2025{\natexlab{a}})\citenamefont {Ruffini}, \citenamefont
  {Prakapenia}, \citenamefont {Quevedo},\ and\ \citenamefont
  {Zhang}}]{Ruffini:2024dwq}%
  \BibitemOpen
  \bibfield  {author} {\bibinfo {author} {\bibfnamefont {Remo}\ \bibnamefont
  {Ruffini}}, \bibinfo {author} {\bibfnamefont {Mikalai}\ \bibnamefont
  {Prakapenia}}, \bibinfo {author} {\bibfnamefont {Hernando}\ \bibnamefont
  {Quevedo}}, \ and\ \bibinfo {author} {\bibfnamefont {Shurui}\ \bibnamefont
  {Zhang}},\ }\bibfield  {title} {\enquote {\bibinfo {title} {{Single versus
  the Repetitive Penrose Process in a Kerr Black Hole}},}\ }\href {\doibase
  10.1103/PhysRevLett.134.081403} {\bibfield  {journal} {\bibinfo  {journal}
  {Phys. Rev. Lett.}\ }\textbf {\bibinfo {volume} {134}},\ \bibinfo {pages}
  {081403} (\bibinfo {year} {2025}{\natexlab{a}})},\ \Eprint
  {http://arxiv.org/abs/2405.08229} {arXiv:2405.08229 [gr-qc]} \BibitemShut
  {NoStop}%
\bibitem [{\citenamefont {Ruffini}\ \emph
  {et~al.}(2025{\natexlab{b}})\citenamefont {Ruffini}, \citenamefont {Bianco},
  \citenamefont {Prakapenia}, \citenamefont {Quevedo}, \citenamefont {Rueda},\
  and\ \citenamefont {Zhang}}]{Ruffini:2024irc}%
  \BibitemOpen
  \bibfield  {author} {\bibinfo {author} {\bibfnamefont {R.}~\bibnamefont
  {Ruffini}}, \bibinfo {author} {\bibfnamefont {C.~L.}\ \bibnamefont {Bianco}},
  \bibinfo {author} {\bibfnamefont {M.}~\bibnamefont {Prakapenia}}, \bibinfo
  {author} {\bibfnamefont {H.}~\bibnamefont {Quevedo}}, \bibinfo {author}
  {\bibfnamefont {J.~A.}\ \bibnamefont {Rueda}}, \ and\ \bibinfo {author}
  {\bibfnamefont {S.~R.}\ \bibnamefont {Zhang}},\ }\bibfield  {title} {\enquote
  {\bibinfo {title} {{Role of the irreducible mass in repetitive Penrose energy
  extraction processes in a Kerr black hole}},}\ }\href {\doibase
  10.1103/PhysRevResearch.7.013203} {\bibfield  {journal} {\bibinfo  {journal}
  {Phys. Rev. Res.}\ }\textbf {\bibinfo {volume} {7}},\ \bibinfo {pages}
  {013203} (\bibinfo {year} {2025}{\natexlab{b}})},\ \Eprint
  {http://arxiv.org/abs/2405.10459} {arXiv:2405.10459 [gr-qc]} \BibitemShut
  {NoStop}%
\bibitem [{\citenamefont {Brown}\ \emph {et~al.}(2024)\citenamefont {Brown},
  \citenamefont {Iliesiu}, \citenamefont {Penington},\ and\ \citenamefont
  {Usatyuk}}]{Brown:2024ajk}%
  \BibitemOpen
  \bibfield  {author} {\bibinfo {author} {\bibfnamefont {Adam~R.}\ \bibnamefont
  {Brown}}, \bibinfo {author} {\bibfnamefont {Luca~V.}\ \bibnamefont
  {Iliesiu}}, \bibinfo {author} {\bibfnamefont {Geoff}\ \bibnamefont
  {Penington}}, \ and\ \bibinfo {author} {\bibfnamefont {Mykhaylo}\
  \bibnamefont {Usatyuk}},\ }\bibfield  {title} {\enquote {\bibinfo {title}
  {{The evaporation of charged black holes}},}\ }\href@noop {} {\  (\bibinfo
  {year} {2024})},\ \Eprint {http://arxiv.org/abs/2411.03447} {arXiv:2411.03447
  [hep-th]} \BibitemShut {NoStop}%
\bibitem [{\citenamefont {Zaslavskii}(2024)}]{Zaslavskii:2024zgh}%
  \BibitemOpen
  \bibfield  {author} {\bibinfo {author} {\bibfnamefont {O.~B.}\ \bibnamefont
  {Zaslavskii}},\ }\bibfield  {title} {\enquote {\bibinfo {title} {{General
  properties of the electric Penrose process}},}\ }\href {\doibase
  10.1103/PhysRevD.109.124053} {\bibfield  {journal} {\bibinfo  {journal}
  {Phys. Rev. D}\ }\textbf {\bibinfo {volume} {109}},\ \bibinfo {pages}
  {124053} (\bibinfo {year} {2024})},\ \Eprint
  {http://arxiv.org/abs/2403.12879} {arXiv:2403.12879 [gr-qc]} \BibitemShut
  {NoStop}%
\bibitem [{\citenamefont {Ruffini}\ and\ \citenamefont
  {Wheeler}(1971)}]{Ruffini:1971bza}%
  \BibitemOpen
  \bibfield  {author} {\bibinfo {author} {\bibfnamefont {Remo}\ \bibnamefont
  {Ruffini}}\ and\ \bibinfo {author} {\bibfnamefont {John~A.}\ \bibnamefont
  {Wheeler}},\ }\bibfield  {title} {\enquote {\bibinfo {title} {{Introducing
  the black hole}},}\ }\href {\doibase 10.1063/1.3022513} {\bibfield  {journal}
  {\bibinfo  {journal} {Phys. Today}\ }\textbf {\bibinfo {volume} {24}},\
  \bibinfo {pages} {30} (\bibinfo {year} {1971})}\BibitemShut {NoStop}%
\bibitem [{\citenamefont {Bravetti}\ \emph {et~al.}(2016)\citenamefont
  {Bravetti}, \citenamefont {Gruber},\ and\ \citenamefont
  {Lopez-Monsalvo}}]{Bravetti:2015xsp}%
  \BibitemOpen
  \bibfield  {author} {\bibinfo {author} {\bibfnamefont {Alessandro}\
  \bibnamefont {Bravetti}}, \bibinfo {author} {\bibfnamefont {Christine}\
  \bibnamefont {Gruber}}, \ and\ \bibinfo {author} {\bibfnamefont {Cesar~S.}\
  \bibnamefont {Lopez-Monsalvo}},\ }\bibfield  {title} {\enquote {\bibinfo
  {title} {{Thermodynamic optimization of a Penrose process: An
  engineers{\textquoteright} approach to black hole thermodynamics}},}\ }\href
  {\doibase 10.1103/PhysRevD.93.064070} {\bibfield  {journal} {\bibinfo
  {journal} {Phys. Rev. D}\ }\textbf {\bibinfo {volume} {93}},\ \bibinfo
  {pages} {064070} (\bibinfo {year} {2016})},\ \Eprint
  {http://arxiv.org/abs/1511.06801} {arXiv:1511.06801 [gr-qc]} \BibitemShut
  {NoStop}%
\end{thebibliography}%

\end{document}